\documentclass[a4paper,11pt,reqno]{amsart}
\usepackage{amssymb,latexsym,times}
\title[Non-linear parent action and dual N=1 D=4 supergravity]
{Non-linear parent action and dual N=1 D=4 supergravity}

\author[A.J. Nurmagambetov]{Alexei J. Nurmagambetov}

\address{A.I. Akhiezer Institute for Theoretical Physics \newline
\indent NSC ``Kharkov Institute of Physics and Technology''
\newline \indent 1 Akademicheskaya St.
\newline \indent Kharkov, 61108, Ukraine}

\thanks{e-mail: ajn@kipt.kharkov.ua}
\begin{document}

\maketitle

\begin{abstract}
We give N=1 supersymmetric extension of D=4 dual gravity non-linear action proposed in \cite{bh08}.
The dual supergravity action and symmetries of the model, but supersymmetry, are realized in a local way.
\end{abstract}

\section{Introduction}

Recently, in \cite{bh08}, it was proposed a new type of the dual gravity non-linear action. Apparent advantage of the approach is locality of the action as well as of gauge symmetries of the theory. Recall, that previously proposed dual gravity non-linear actions were realized within a non-local approach, when the action and gauge symmetries of the model contain non-local parts
\cite{ajn04,ajn06} (see also \cite{ellwanger06}). At the same time it is worth mentioning that the dynamics of the dual to graviton field is described, after transition to new variables \cite{ajn06}, by a local equation of motion.

Non-locality of the approach of \cite{ajn04,ajn06} is a direct consequence of non-locality in duality relations between graviton and its dual partner. The duality relation is a first order in space-time derivatives expression which contains both fields dynamics. In dependence on the basic field choice the duality relation results either in the graviton dynamical equation or in the equation of motion of its dual partner. The action from which the duality relation comes was realized in \cite{ajn04,ajn06} in a duality-symmetric manner, that is, it includes, in its final form\footnote{This form of the action is equivalent to that of \cite{ajn04}.}, kinetic terms (and terms of interactions) of the original and of the dual to them fields. A most fascinating example is given by the action of the bosonic subsector of the completely duality-symmetric D=11 supergravity \cite{ajn04}, which extends the construction of \cite{bbs97}. The duality-symmetric content of fields entering the action reproduces the corresponding structure of low levels generators (up to the level three) in the proposed infinitely-dimensional hidden symmetry algebra of M-theory \cite{west&}. This algebra predicts the dual to graviton field which should be incorporated into the  effective dynamics of M-theory. It gives another reason to invoke non-locality in view of numerous no-go theorems \cite{bb&} which claim on impossibility to construct the dual non-linear spin-2 field theory action by local deformations of free theory.

The abovementioned construction is realized within the second order formalism, when the basic (and unique) variables of the problem are vielbeins, while the spin-connection is expressed through the vielbeins and their derivatives.
In contrast to \cite{ajn04,ajn06}, the formulation of \cite{bh08} is based on the first order reformulation of gravitational action that results in a number of important changes. The principle change consists in dualisation of the spin-connection and not of the vielbein, as it was done in \cite{ajn04,ajn06}. Dualising the spin-connection is easy to realize in the linearized limit \cite{bch&} (see also \cite{nieto}, \cite{hull00} for early papers), which turns out to be generalized to the non-linear case \cite{bh08}. The non-linear first order action of \cite{bh08} is constructed out the dual spin-connection, extended, in a specific way, with the dual to the vielbein ``field strength''. It also contains the original vielbeins, entering the action through a Chern-Simons combination with the dual connection. Another distinctive property of the construction is its straightforward generalization to the case of interacting with matter fields. The formulation of \cite{bh08} still remains local in the case. Recall that the second order formalism is ``sensitive'' to the presence of matter fields, when dualisation requires, even in the linearized limit,  introducing the non-locality in the duality relations \cite{ajn08}, \cite{bk&08}.

We have noticed the relation between the hidden symmetry algebra of M-theory \cite{west&} and the structure of the duality-symmetric action \cite{ajn04,ajn06}. It is also worth mentioning the r\^{o}le of supersymmetry in this respect, because the starting point in searching for the M-theory hidden symmetry structure was D=11 supergravity and the structure of its hidden symmetries arising upon the toroidal dimensional reduction \cite{west&}. The low level generators of the proposed hidden symmetry algebra, known now as $E_{11}$, are in one-to-one correspondence with fields which cast N=1 supersymmetric spin-2 multiplet in eleven space-time dimensions, including the dual to graviton and to a third rank antisymmetric tensor fields\footnote{The structure of the first order parent action within $E_{11}$ perspective was discussed in \cite{bh08}.}. Therefore, we have to get the supersymmetric formulation of the dual to graviton field theory in the end. Searching for a realization of this task is the main motivation of the paper.

The paper is organized as follows. In the next section we make preliminary steps to obtain N=1 supersymmetric extension of the non-linear parent action of D=4 dual gravity \cite{bh08}. Doing that, we make a special accent on the supersymmetry transformations of spin-connection. Usually these transformations are ignored since their explicit form does not needed to prove the invariance of a supergravity action under the local supersymmetry transformations (in the so-called 1.5 formalism\footnote{See e.g. \cite{vNPR} for details, and \cite{js00} for discussing the relation of 1.5 formalism to the first order supergravity formulation.}). However, the explicit form of the spin-connection supersymmetry transformation essentially simplifies extracting the corresponding transformation of the dual to the vierbein field. The supersymmetric extension of the dual gravity parent action is given, in its final form, in Section 3. Summary of the results is contained in Conclusions. Our notation and conventions are listed in Appendix.

\section{Prerequisites of the construction}

We begin with the standard action for D=4 N=1 supergravity \cite{VS73,FFN76,DZ76}
\begin{equation}\label{EH}
S_{EH+RS}=-\frac{1}{4k^2}\int \, d^4 x \, eR(\omega)-\frac{i}{2}\int \, d^4 x \, e\bar{\psi}_m\gamma^{mnk}D_{n}(\omega)\psi_k,
\end{equation}
where $e=\det{e^a_m}$ is the determinant of the vierbein, the curvature tensor is defined by
\begin{equation}\label{Rdef}
{R_{mn}}^{ab}(\omega)=\partial_m {\omega_n}^{ab}+{\omega_m}^{ac}{\omega_{nc}}^b-(m\leftrightarrow n),
\end{equation}
and $R(\omega)=e^{m}_a e^{n}_b {R_{mn}}^{ab}(\omega)$. The spin-connection ${\omega_m}^{ab}=-{\omega_m}^{ba}$ and the vierbein are independent fields that corresponds to the first order formulation.

Action (\ref{EH}) is invariant under space-time diffeomorphisms, local Lorentz transformations which act on the flat-type indices, and local supersymmetry transformations
\begin{equation}\label{SUSYst}
\delta_{\epsilon}e_{m}^a=-ik\bar{\epsilon}\gamma^a \psi_m,\qquad \delta_{\epsilon} \psi_m=\frac{1}{k}D_m(\omega)\epsilon,
\end{equation}
with a parameter $\epsilon(x)$. The supersymmetry transformation of the connection \cite{DZ76} will also be important for our aims. In our notation it reads
\begin{equation}\label{susycon}
\delta_{\epsilon}{\omega_a}^{bc}=\frac{1}{4}k\varepsilon^{bcde}\bar{\epsilon}\gamma_5 \left[ \gamma_e \Psi_{da}(\omega)-\gamma_a\Psi_{ed}(\omega)+\gamma_d\Psi_{ae}(\omega)\right],
\end{equation}
where we have introduced $\Psi_{ab}=e^m_a e^n_b(D_m(\omega)\psi_n-D_n(\omega)\psi_m)$.

Our next step is to transform the action into a form which would be convenient for dualisation. We note to this end that (\ref{EH}), up to irrelevant boundary terms, may be rewritten in the following form
\[
S=-\frac{1}{4k^2}\int \, d^4 x e\left[ {\omega_a}^{ab}{\omega_{cb}}^c-\omega_{abc}\omega^{bca}+2{C_{ab,}}^a{\omega_c}^{bc}+C_{ab,c}\omega^{cab} \right ]
\]
\begin{equation}\label{SGew}
-\frac{i}{2}\int \, d^4 x \, e\bar{\psi}_m\gamma^{mnk}D_{n}(\omega)\psi_k ,
\end{equation}
with ${C_{ab,}}^c=2\partial_{[a}e^c_{b]}$. This action, or, more precisely, the action which apparently relates to (\ref{SGew})
\[
S=-\frac{1}{4k^2}\int \, d^4 x e [ e^m_a e^n_c {\omega_m}^{ab}{\omega_{nb}}^c-e^m_a e^n_b\omega_{mbc}\omega^{nca}+2e^m_a e^n_b{C_{mn,}}^a{\omega_c}^{bc}
\]
\begin{equation}\label{SGew1}
+e^m_a e^n_b e_k^c C_{mn,c}\omega^{kab}  ]-\frac{i}{2}\int \, d^4 x \, e\bar{\psi}_m\gamma^{mnk}D_{n}(\omega)\psi_k,
\end{equation}
${C_{mn,}}^a=2\partial_{[m}e^a_{n]}$, is invariant under the local supersymmetry transformations (\ref{SUSYst}), (\ref{susycon}).

Next, we change the connection to the following variables \cite{bch&}
\begin{equation}\label{Y}
\omega_{abc}=Y_{bc,a}+\eta_{a[b}{Y_{c]d,}}^d, \qquad Y_{ab,c}=-Y_{ba,c}.
\end{equation}
It results in
\[
S=-\frac{1}{4k^2}\int \, d^4 x e\left[ e^m_a e^n_b C_{mn,c}Y^{ab,c}+Y_{ab,c}Y^{ac,b}-\frac{1}{2}{Y_{ab,}}^b {Y^{ac,}}_c
\right ] \]
\begin{equation}\label{CY}
-\frac{i}{2}\int \, d^4 x \, e\bar{\psi}_m\gamma^{mnk}D_{n}(Y)\psi_k,
\end{equation}
where
\begin{equation}\label{DY}
D_m(Y)\psi_n=\left( \partial_m+\frac{1}{4}{Y^{ab,}}_m \gamma_{ab}+\frac{1}{4}e^a_m {Y^{bc,}}_c \gamma_{ab} \right )\psi_n.
\end{equation}
The supersymmetry transformations which leave (\ref{CY}) invariant are
\begin{equation}\label{CYsusytr}
\delta_{\epsilon}e_{m}^a=-ik\bar{\epsilon}\gamma^a \psi_m,\qquad \delta_{\epsilon} \psi_m=\frac{1}{k}D_m(Y)\epsilon,
\end{equation}
\[
\delta_{\epsilon} Y_{ab,c}=\frac{1}{4}k{\varepsilon_{ab}}^{fd}\bar{\epsilon}\gamma_5 \left[ \gamma_d \Psi_{fc}(Y)-\gamma_c \Psi_{df}(Y)+\gamma_f \Psi_{cd}(Y)\right ]
\]
\begin{equation}\label{Ysusytr}
+\frac{3}{2}k\eta_{c[a}{\varepsilon^e}_{b]fd}\bar{\epsilon}\gamma_5\gamma^d {\Psi^f}_e(Y).
\end{equation}

Finally, we write down the action (\ref{CY}) in terms of the dual variables
\begin{equation}\label{U}
Y^{ab,c}=\frac{1}{2}\varepsilon^{abdf}{\mathcal{Y}_{df,}}^c,
\end{equation}
after that it becomes
\[
S=-\frac{1}{4k^2}\int \, d^4 x e\left[ \frac{1}{2}e^m_a e^n_b C_{mn,c}\varepsilon^{abdf}{\mathcal{Y}_{df,}}^{c}-\frac{1}{4}\mathcal{Y}_{ab,c}\mathcal{Y}^{ab,c}-\frac{1}{2}\mathcal{Y}_{ab,c}\mathcal{Y}^{ac,b}+{\mathcal{Y}_{ab,}}^b {\mathcal{Y}^{ac,}}_c
\right ]
\]
\begin{equation}\label{CU}
-\frac{i}{2}\int \, d^4 x \, e\bar{\psi}_m\gamma^{mnk}D_{n}(\mathcal{Y})\psi_k .
\end{equation}

The covariant derivative acting on gravitino is
\begin{equation}\label{DUpsi}
D_m(\mathcal{Y})\psi_n=\left(\partial_m+\frac{1}{8}\varepsilon^{abcd}e^e_m \gamma_{ab}\mathcal{Y}_{cd,e}+\frac{1}{8}e^a_m \varepsilon^{bcde}\gamma_{ab}\mathcal{Y}_{de,c}\right )\psi_n,
\end{equation}
and (\ref{CU}) is invariant under the following local susy transformations
\[
\delta_{\epsilon}e_{m}^a=-ik\bar{\epsilon}\gamma^a \psi_m,\quad \delta_{\epsilon} \psi_m=\frac{1}{k}D_m(\mathcal{Y})\epsilon,
\]
\begin{equation}\label{CUsusytr}
\delta_{\epsilon}{\mathcal{Y}_{ab,c}}=k\bar{\epsilon}\gamma_5\left( \gamma_c \Psi_{ba}(\mathcal{Y})+2\gamma_b \Psi_{ac}(\mathcal{Y})+2\gamma_a \Psi_{cb}(\mathcal{Y})\right ),
\end{equation}
with the covariant gravitino curl $\Psi_{ab}(\mathcal{Y})$
\[
\Psi_{ba}(\mathcal{Y})=\partial_b \psi_a+\frac{i}{8}\gamma_5 \gamma^{fg}\mathcal{Y}_{fg,b}\psi_a+
\frac{1}{4}\varepsilon_{fgeb}\mathcal{Y}^{fg,e}\psi_a-\frac{3i}{4}\gamma_5\gamma^{eg}\mathcal{Y}_{bg,e}\psi_a
\]
\begin{equation}\label{PsiY}
-(a\leftrightarrow b).
\end{equation}

\section{The parent action}

After these preparations we are now ready to construct the non-linear parent action of dual N=1 D=4 supergravity. Following \cite{bh08} we extend the dual connection $\mathcal{Y}_{ab,c}$ with the curl of a new field. The dual spin-connection becomes
\begin{equation}\label{calF}
\mathcal{F}_{ab,c}=\mathcal{Y}_{ab,c}+F_{ab,c},
\end{equation}
where
\begin{equation}\label{F}
{F_{ab}}^{,c}=2\partial_{[a}{\mathcal{C}_{b]}}^{,c}.
\end{equation}
The field $\mathcal{C}_{a,b}$ is the dual to graviton field.

Substituting the dual spin-connection (\ref{calF}) into (\ref{CU}) we arrive at
\[
S=-\frac{1}{4k^2}\int \, d^4 x e\left[ \frac{1}{2}e^m_a e^n_b C_{mn,c}\varepsilon^{abdf}{\mathcal{Y}_{df,}}^{c}-\frac{1}{4}\mathcal{F}_{ab,c}\mathcal{F}^{ab,c}-\frac{1}{2}\mathcal{F}_{ab,c}\mathcal{F}^{ac,b}+{\mathcal{F}_{ab,}}^b {\mathcal{F}^{ac,}}_c
\right ]
\]
\begin{equation}\label{CF}
-\frac{i}{2}\int \, d^4 x \, e\bar{\psi}_m\gamma^{mnk}D_{n}(\mathcal{F})\psi_k .
\end{equation}
This is the supersymmetric extension of the action constructed in \cite{bh08}. The covariant derivative acting on  gravitino is
\begin{equation}\label{DFpsi}
D_m(\mathcal{F})\psi_n=\left(\partial_m+\frac{1}{8}\varepsilon^{abcd}e^e_m \gamma_{ab}\mathcal{F}_{cd,e}+\frac{1}{8}e^a_m \varepsilon^{bcde}\gamma_{ab}\mathcal{F}_{de,c}\right )\psi_n,
\end{equation}
and we have omitted in (\ref{CF}) the total derivative term $de^a\wedge d\mathcal{C}^b \eta_{ab}$.

The action we get possesses the same symmetries as its non-supersymmetric predecessor. It is invariant under the local diffeomorphisms, local Lorentz transformations and special shift symmetry
\begin{equation}\label{shiftF}
\delta{\mathcal{Y}_{ab}}^{,c}=2\partial_{[a}{\Sigma_{b]}}^{,c}, \qquad \delta{\mathcal{C}_{a}}^{,b}=-{\Sigma_{a}}^{,b},
\end{equation}
which leaves $\mathcal{F}_{ab,c}$ invariant. The analysis of these symmetries is the same as it has been done in \cite{bh08}, so we refer the reader to this reference for details.

The rest of the symmetries of (\ref{CF}) is the local supersymmetry to consideration of which we now turn. The supersymmetry transformations which leave the action invariant are
\[
\delta_{\epsilon}e_{m}^a=-ik\bar{\epsilon}\gamma^a \psi_m,\quad \delta_{\epsilon} \psi_m=\frac{1}{k}D_m(\mathcal{F})\epsilon,
\]
\begin{equation}\label{CFsusytr}
\delta_{\epsilon}{\mathcal{F}_{ab,c}}=k\bar{\epsilon}\gamma_5\left( \gamma_c \Psi_{ba}(\mathcal{F})+2\gamma_b \Psi_{ac}(\mathcal{F})+2\gamma_a \Psi_{cb}(\mathcal{F})\right ).
\end{equation}
Since we have known the transformation law of $\mathcal{Y}_{ab,c}$, eq. (\ref{CUsusytr}), the local Susy transformation of the dual ``field strength'' is read off (\ref{CFsusytr}). It results in
\begin{equation}\label{mathCsusytr}
\delta_{\epsilon}\partial_{[a}\mathcal{C}_{b,]c}=\frac{1}{2}k\bar{\epsilon}\gamma_5 \left[ \gamma_c \Phi_{ba}(\partial \mathcal{C})+2\gamma_b \Phi_{ac}(\partial\mathcal{C})+2\gamma_a\Phi_{cb}(\partial\mathcal{C})\right ],
\end{equation}
where
\begin{equation}\label{Phidef}
\Phi_{ba}(\partial \mathcal{C})=\frac{i}{8}\gamma_5 \gamma^{fg}\partial_f \mathcal{C}_{g,b}\psi_a+\frac{1}{4}\varepsilon_{fgeb}\partial^f\mathcal{C}^{g,e}\psi_a-\frac{3i}{4}\gamma^{eg}\partial_b\mathcal{C}_{g,e}\psi_a-(a\leftrightarrow b).
\end{equation}

Let us turn to the discussion of (\ref{mathCsusytr}). Note first that (\ref{CFsusytr}) is similar to a standard form of the spin-connection supersymmetry transformation: the variation of the connection is expressed through a function depending on the connection (see  (\ref{susycon}), (\ref{Ysusytr}), (\ref{CUsusytr})). The dual to graviton field enters the generalized connection $\mathcal{F}_{ab,c}$ through the ``field strength'' $F^a=dC^a$. Its transformation law is a part of the supersymmetry transformations (\ref{CFsusytr}), hence (\ref{mathCsusytr}) has the standard, if we treat it as a spin-connection, form. On the other hand, one may notice that ${C_m}^{,a}$ is the dual to $e^a_m$ field. The vierbein possesses its own transformation under the local Susy, eq. (\ref{SUSYst}). The same may be expected for ${C_m}^{,a}$, so the problem is how to read off the Susy transformation of `bare' ${C_m}^{,a}$ from (\ref{mathCsusytr}).

We have not a good recipe to solve this problem. The point is that the dual field transforms under the local Lorentz as \cite{bh08}
\begin{equation}\label{CLL}
\delta{\mathcal{C}_a}^{,b}={\tilde{\Lambda}_a}^{~b}+\dots,
\end{equation}
with the dual parameter ${\tilde{\Lambda}_a}^{~b}=-{\tilde{\Lambda}^b}_{~a}$. We can fix ${\tilde{\Lambda}_a}^{~b}$ to remove the antisymmetric part of ${\mathcal{C}_a}^{,b}$, i.e. ${\mathcal{C}_{[a}}^{,b]}=0$. Then, eqs. (\ref{mathCsusytr}), (\ref{Phidef}) get transformed into
\begin{equation}\label{Cgfsusytr}
\delta_{\epsilon}\partial_{[a}\mathcal{C}_{b],c}|_{\rm{gauge~ fixed}}=-\frac{i}{8}k\bar{\epsilon}\mathcal{M}_{cba}^{efg}\gamma_e (\gamma^{ij}\partial_i \mathcal{C}_{j,f})\psi_g,
\end{equation}
where we have introduced
\[
\mathcal{M}_{cba}^{efg}=\delta^e_c\delta^{fg}_{ba}+2\delta^e_b\delta^{fg}_{ac}+2\delta^e_a\delta^{fg}_{cb}, \qquad \delta^{fg}_{ba}\equiv \delta^f_{[b}\delta^g_{a]}.
\]
Clearly, eq. (\ref{Cgfsusytr}) sets a barrier to extract the supersymmetry transformation of ${C_m}^{,a}$ in a local way.

\section{Conclusions}

To summarize, we have constructed the supersymmetric extension of the parent action of dual D=4 gravity proposed in \cite{bh08}. This procedure does not break the locality of the action, its symmetries (excluding supersymmetry, cf. (\ref{mathCsusytr}), (\ref{Cgfsusytr}), and see the discussion below) and equations of motion. Constructing the action we followed the same routine of field transformations which was realized in \cite{bch&}, \cite{bh08}, getting started with classical action of D=4 simple supergravity. Using the explicit form of the spin-connection supersymmetry transformations, usually ignored, essentially simplified the derivation of supersymmetry rules after dualisation.

Before introducing the dual to graviton field all the Susy transformations are local. Once the dual field is taken into account non-locality appears through the dual field strength supersymmetry transformation. We can not exclude that the appropriate local transformation of the field will be found in the end, however its explicit form remains obscure for a while.

On the other hand, we can not also definitely exclude the absence of the dual field Susy transformation in a local form. A reason for that follows from the previous experience in the construction of hidden symmetry algebra upon the toroidal dimensional reduction of N=1 D=4 supergravity to one dimension \cite{nicolai91}. We remind that fermions are needed to enlarge the hidden symmetry algebra in one dimension to a hyperbolic Kac-Moody algebra. The enlargement is realized due to an interplay between two types of transformations, the Ehlers transformations and the Matzner-Misner transformations (see \cite{nicolai91} for details). These transformations are non-local when they act on the dual bosonic fields. Supersymmetry has to be a part of the hidden symmetry structure, and, perhaps, acts also non-locally on the dual to graviton field. Anyhow, this point requires additional studies.

Another problem is to obtain a modification of supersymmetry transformations of parent actions in the case of extended or higher-dimensional supergravities, when antisymmetric gauge fields are taken into account. It would be interesting, in such a perspective, to realize an applicability of the approach of \cite{bh08} to duality-symmetric theories, in particular, to the description, in a manifestly covariant way, of the self-dual part of D=10 type IIB supergravity action.

\vspace{0.3cm}

\noindent {\bf Acknowledgments.} The author is grateful to A.N. Petrash for help in verifying the supergravity transformations. Work supported in part by the INTAS grant \# 05-08-7928, and the joint NASU--RFFR grant \# 38/50-2008.

\section*{A. Notation and conventions}

Our notation is as follows: indices from the beginning of the Latin alphabet are vector flat indices, whose of the middle of the Latin alphabet are curve. Spinor indices are hidden.

The Levi-Civita tensor is chosen to be
\[
\varepsilon_{abcd}\varepsilon^{abcd}=-4!,
\]
we use the mostly minus signature with $\eta_{ab}={\rm diag} (+,-,-,-)$.

Torsion and curvature tensors are defined by
\[
T^a=de^a+{\omega^a}_b \wedge e^b,
\]
\[
R^{ab}=d\omega^{ab}+\omega^{ac}\wedge{\omega_c}^b,
\]
that, in the coordinate basis, corresponds to
\[
{T_{mn}}^a=\partial_m e^a_n+{\omega_m}^{ab}e_{nb}-(m\leftrightarrow n)
\]
and to eq. (\ref{Rdef}). The external derivative $d$ acts from the left.

For $\gamma$-matrices we use
\[
\gamma^{ab}=\frac{1}{2}[\gamma^a,\gamma^b],\qquad \gamma^{abc}=\frac{1}{2}\{\gamma^a,\gamma^{bc}\}.
\]
We define
\[
\gamma_5=i\gamma^0\gamma^1\gamma^2\gamma^3,
\]
then
\[
\gamma^{abcd}=-i\gamma_5 \varepsilon^{abcd},\qquad \gamma^{abc}=-i\varepsilon^{abcd}\gamma_5\gamma_d.
\]

Spinors are Majorana, the covariant derivative is defined by
\[
D_m\psi_n=\left( \partial_m+\frac{1}{4}\omega_{mab}\gamma^{ab}\right)\psi_n.
\]

\end{document}